# THE META-STRUCTURES PROJECT

Gianfranco Minati
Italian Systems Society, Milan, Italy
gianfranco.minati@AIRS.it

**INTRODUCTION**

In this paper we introduce four research projects based on the concept of meta-structure. The first one relates to the theoretical basis and research to identify meta-structures and their properties in simulated or natural collective behaviours. Among the several objectives of the project listed below, one is the implementation of software for simulation and search for meta-structures in simulated collective behaviour to be released *free downloadable*. This software is expected to be resources for all other projects. This first project relates to the theory of meta-structures for modelling the emergence of collective phenomena based on variable structures between mesoscopic variables and as an initial contribution towards a general theory of emergence.

The second project relates to transformation, to be possibly modelled as being due to meta-structural properties, of biological survival needs of living matter into behaviour, as well as phases of matter, such as *living*, and how they are related to meta-structural properties.

The third project relates to the application of meta-structures for modelling constraints in social systems considered as meta-structures able to influence processes of emergence of acquired properties. One aspect of this project relates to revisiting the concept of Marx's superstructure considered now as processes of emergence.

The fourth project relates to the structuring of living space designed by architects as meta-structuring able to influence the emergence of social behavioural properties. Architecture is intended as the self-design of meta-structures (boundary conditions) to induce and maintain, through structuring, emergent properties in human social systems.

The first project relates to the possibility of identifying suitable meta-structural properties to model the acquisition of properties in collective behaviours. In a similar way, the second project relates to the identification of meta-structural properties suitable for modelling processes of emergence occurring in living matter. The other two projects relate to considering design constraints, as for social systems and the architectural structuring of space, as meta-structures influencing the processes of acquisition of emergent properties in social systems.

This project is in collaboration with my *current research associates,* Arne Collen, Saybrook Graduate School and Research Center, Eliano Pessa, University of Pavia, Giordano Bruno, University "La Sapienza" of Rome, Giuseppe Vitiello, University of Salerno, Ignazio Licata, Institute for Basic Research, Larry A. Magliocca, Ohio State University, Salvatore Di Gregorio, University of Calabria and Valerio Di Battista, Polytechnic of Milan.

This project is also in collaboration with colleagues of the Italian Systems Society www.AIRS.it as outlined in the proceedings of its national conferences (Minati, Abram and Pessa, 2008; Minati, Pessa and Abram, 2006; Minati and Pessa, 2002; Minati, 1998).

## 1. THE THEORY OF META-STRUCTURES FOR MODELLING THE EMERGENCE OF COLLECTIVE PHENOMENA AS AN INITIAL CONTRIBUTION TOWARDS A GENERAL THEORY OF EMERGENCE.

The theory, introduced in 2008 (Minati, 2008a; 2008b; 2008c; 2008d; 2008e), is based on considering properties of dynamical sets of mesoscopic variables abductively chosen by the observer to model processes of acquisition of emergent properties in collective phenomena. Contrary to microscopic and macroscopic variables, mesoscopic variables relate to *aggregates* of microscopic but not macroscopic values.

State variables are assumed to completely describe the state of the system under study. For instance:

- Microscopic state variables are assumed to describe the behaviour of a system through the behaviour of single, interacting components, as in the solar system and in simple mechanical systems.
- Macroscopic state variables are assumed to describe the behaviour of a system through macroscopic variables representing, for instance, temperature, pressure and density of a thermodynamic system.

**1.1 Meta-structures for modelling the emergence of collective phenomena**

With reference to collective behaviours established by interacting agents, microscopic variables might be the distance between single agents or their specific speed and direction over time, whereas macroscopic variables may be the density, volume or surface of the system over time as temperature and pressure are considered for thermodynamic systems:

- Mesoscopic state variables may relate, for instance, to the number of elements having the *same* value (the observer considers values as *equal* when they lie *within a range of values*) taken by some specific variables such as elements having the *same* distance from their nearest neighbours, elements having the *same* speed or elements having the *same* direction over time. We specify that mesoscopic variables consider clusters composed of elements lying at the same distance. Different clusters correspond to different distances of the composing elements. For instance, $n$ different elements constituting a mesoscopic variable may be clustered into groups having the *same* distance, say $n_1$ are at distance $d_1$, $n_2$ are at distance $d_2$, etc., in such a way that $n_1 + n_2 + \ldots + n_s = n$, with $n_i > k$ where $k$ is the *granularity* of the mesoscopic level of description. The same applies to speed and direction. For instance,
    - *Mx* is the number of elements having their maximum distance at a given point in time;
    - *Mn* is the number of elements having their minimum distance at a given point in time;
    - $N_1(t)$ is the number of elements having the *same* distance from their nearest neighbour at a given point in time;
    - $N_2(t)$ is the number of elements having the *same* speed at a given point in time;
    - $N_3(t)$ is the number of elements having the *same* direction at a given point in time.
  
  This level of description is suitable for considering the percentage of elements having the *same* property and percentage of time spent by the elements possessing this property to identify *mesoscopic ergodicity* at this level of description of the system.

- Meta-elements. Time-ordered sets of values in a discrete temporal representation, specifying mesoscopic state variables are considered as *meta-elements*. While mesoscopic state variables take a value, for instance, of the number of elements having the same property over time, meta-elements are sets of corresponding values assumed by the property, such as distance, speed and direction, considered over time. Examples of meta-elements at each instant $t$ are given by considering over time values specifying mesoscopic state variables, including:

    - *MD(t)* is the value of the maximum distance at time $t$;
    - *MN(t)* is the value of the minimum distance at time $t$;
    - *Dn(t)* is the value of the distance considered suitable to specify the mesoscopic variable at time $t$, e.g., max or min or weighted average among those corresponding to different simultaneous clusters;

- *Sp(t)* is the value of the speed considered suitable for specifying the mesoscopic variable at time *t*, e.g., max or min or weighted average among those corresponding to different simultaneous clusters;
- *Dir(t)* is the value of the direction considered suitable for specifying the mesoscopic variable at time *t*, e.g., max or min or weighted average among those corresponding to different simultaneous clusters.

  o Meta-structure. The properties of meta-elements such as interpolating functions, statistical, levels of ergodicity -*meta-structural ergodicity*-, and possible relationships between them in an *N-space*, suitable to model a kind of *entropy of correlations*, establish the meta-structure. While structure relates to interacting elements, meta-structures relate to mesoscopic state variables and their parameters. Meta-elements may be considered in a phase space where each mesoscopic state variable of the system is associated with a coordinate axis. System behaviour can be represented as the motion of a point along a trajectory in this space. In this conceptual framework we may apply usual approaches considered in physics, like existence of periodic or strange attractors, and fixed points.

Meta-structures have been introduced to model, at a mesoscopic level, collective behaviour and processes of acquisition of emergent properties in collective phenomena.

The approach has been introduced after a suitable reformulation of the concepts of *systemic coherence* (Minati, 2008a, pp. 39-40), self-organisation (Minati, 2008a, pp. 62-64) and emergence (Minati, 2008a, pp. 64-65 and 51-56). In short, this relates to considering processes of self-organisation as *regular* continuous variability of the structure of a process. Variability is stable, i.e., repetitive and foreseeable. Stability of variability corresponds to stability of the acquired property. Emergence relates to continuous processes of self-organisation when variability of the structure is, dynamic, irregular, but *coherent*.

The hypothesis is that through meta-elements and meta-structures, i.e., a meta-level, it is possible to model and manage at a higher and more suitable level of representation, emergent phenomena, i.e., the structural variability through mesoscopic variables and meta-structural properties.

**1.2 Ergodicity**

Different disciplinary definitions of ergodicity have been introduced in the literature, see, for instance, Minati and Pessa (2006, pp. 293-300) for an overview considering physics, theoretical geomorphology, economics, and population dynamics.

The concept may be in short summarised by the following definitions used in theoretical geomorphology (Thorn, 1982):

> "A system is ergodic when the mean of observations of an individual made over time is equal to the mean of observations made of many individuals at a single moment in time over an area." (Thorn, 1988 page 47),

and the *equivalent*

> "If 15% of the population is in a particular state at any moment in time, one can assume that each individual in the population spends 15% of time in that state." (Thorn, 1988, page 48).

In the field of population studie*s* the concept of ergodicity has been introduced mainly to support a technique for reconstructing the past evolution of a population starting from actual data, known as *inverse projection*. This was introduced by Ronald Lee (Lee 1974; 1978; 1985), as a logical inversion of conventional projection techniques.
Reconstruction of population structure over time requires knowledge of its age structure at an initial time. These data are not usually available.

Ordinary projection is a way of passing *from rates to counts*, "say from a sequence of age-specific rates of giving birth and dying, to count over time total births and deaths" (Wachter, 1986). *Inverse projection* is a way of passing *from counts to rates* "from total counts of events like births and deaths to a plausibly responsible sequence of age-specific rate sets chosen from within a model family (from total counts and age-specific rates together, the changing age structure over time is reconstructed)" (Wachter, 1986). In this way it is possible to reconstruct the structure of a population over a specific time-span through the availability of historical series of births and deaths.

This is considered as being related to *ergodic reasoning*, i.e., *trading* the evolution of one variable with another, as for the so-called weak ergodicity theorem (Norton, 1928). This theorem states that different populations subject to identically varying birth-rates would converge to the same age distribution. Because of this, in the long run, birth-rates and not initial states determine what a population will look like (Wachter, 1986; 1997). Dynamical social systems quickly forget (like *initial conditions* in physics) *population age structures*. The *weak ergodicity theorem* states that populations starting from different age structures, having the same initial population size and same dynamics of migration, births, and deaths, will converge to the same age distribution.

**In an *analogous* way we may consider a weak ergodic reasoning based upon considering mesoscopic variables rather than micro or macroscopic ones and specifically sets of values of meta-elements within a given timeframe. The *weakness* is not related to reverse reasoning, but to the dynamic of mesoscopic variables representing the variability of structures.**

The concept of index of *index of ergodicity*, introduced in (Minati and Pessa, 2006, pp. 306-309) is very helpful to detect occurring of processes of emergence of collective behaviour.
Let us consider the behavioural feature $F$ of a system, assumed to be associated with a finite number of different possible states $F_i$.
Let us assume the system is established by a finite number of elements constant over time. The observer may detect that the average percentage of time spent by a single element in state $F_i$ is $y\%_i$ and the average percentage of elements lying in the same state as $x\%_i$. A measure of the *degree of ergodicity* of that particular state can be expressed as:

$$E_i = 1 / [1 + (x\%_i - y\%_i)^2].$$

When $x\%_i = y\%_i$, the index $E_i$ reaches its maximum value of 1 and shows that the system is ergodic. All other values may be used to detect processes of transition associated with the emergence, for instance, of collective behaviours.
When the observer is an integral part of the system under study the index is useful for detecting the occurrence of processes of restructuring, such as phase-transitions, self-organisation and emergence. This is interesting for studying the emergence of collective behaviour, transitions from non-collective to collective behaviour and *vice versa*. Such transitions are associated with variations in the index of ergodicity.
**The index of ergodicity may by applied to mesoscopic variables and meta-structures not only to detect the occurrence of processes of collective behaviour, but, possibly, the specific collective process represented by changes in the index of ergodicity. This is a research topic.**

**1.3 Meta-structural analysis and objectives of the project**

The approach is to be generalised through *meta-structural analysis* (Minati, 2008a, pp. 99-100), having the purpose to identify:
1. Methodologies to make sets of interacting components adopt a specific meta-structural property;
2. Correspondences between meta-structural properties and systemic properties.

Objectives of the projects are:
"
(1) Representation of meta-structures, i.e., properties of meta-elements and the possible, variable relationships between them;
(2) Methodologies for the identification of meta-structures in *general* collective phenomena established by so-called processes of self-organisation;
(3) Identification of relationships between meta-structures and collective phenomena;
(4) Methodologies for applying, 'prescribing', a meta-structure to the behaviour of single agents establishing self-organised systems. It is also a bottom-up approach, based on hypothesizing some meta-structures, simulating interacting particles with such constraints and see what kind of system, if any, is established;
(5) Methodologies for using meta-structures to establish and manage collective phenomena *in general*;
(6) Inquiry into possible *sources* of meta-structures in collective phenomena by introducing theoretical hypotheses based, for instance, on the theory of stability or relationships amongst universal constants and considering possible physical and cognitive characteristics of agents. Another, constructivist, approach can be based upon considering this view as a *cognitive need* of the observer, modelling processes as such because of the cognitive model adopted (I suggest focussing upon this second approach).
(7) Discover how parametric values of meta-structural properties, such as statistical or ergodic, may change over time in correspondence with changes in a system having collective behaviour (e.g., with a changing number of elements)?
(8) Discover whether parametric values of meta-structural properties, such as statistical or ergodic, correspond to specific kinds of collective behaviour (e.g., flocking or swarming)
(9) Explore formal properties related to correspondence between coherence and meta-properties: existence, measurements, and typologies.
(10) Explore the difference between *meta-structural property* and *order parameter*. It is possible to consider attractors in the phase space of meso-variables?
(11) Relationships between quantum models and meta-structural level o descriptions with particular reference to *quantum correlations* (entanglement), see, for instance, (Vedral *et al.*, 1997; Vedral and Plenio, 1998).
(12) Formulation on this basis of initial approaches towards a *General Theory of Emergence of Collective Phenomena*." (Minati, 2008a, pp. 93-94).

## 1.4 Experimental simulation and applications

Simulation software is under development. This software, initially based for convenience on Cellular Automata, will simulate collective behaviour assumed by interacting agents and will make available all positional and dynamic information of agents suitable to compute values assumed by mesoscopic values as freely defined by the observer. The software is initially designed to simulate collective behaviour established by flocks of boids. This software is expected to help search of meta-structures in simulated collective behaviour.
This software will be free downloadable to researchers.
Sophisticated technologies able to extract positional and dynamic information from images of real collective phenomena, like flocks of birds and swarms, may be used in future for searching of meta-structures in real collective phenomena.
Other sophisticated technologies may be also applied for search of meta-structures at microscopic scale.
We mention a field of research (see chapter 2) that could be effectively explored by using meta-structures. It relates to the possibility to represent phases of matter considered in physics by using meta-structures. This possibility may be suitable when considering how variations of ergodicity are useful to represent phase transitions. It is well known that the property of ergodicity is completely lost during a structural change or a *phase transition*.

Solid-liquid-gas transitions are classified as *first-order transitions* because they involve a discontinuous change in density (a discontinuity in the first derivative of the free energy with a thermodynamic variable). *Second-order transitions* (with a discontinuity in the second derivative of the free energy) consist of an internal rearrangement of system structure, occurring *at the same time at all points*. Second-order transitions occur because the conditions for the stable existence of the structure as in the previous phase are no longer active and a new stable structure, replacing the previous one, is established.

A variation in ergodicity occurs between the absence and the birth of coordination. Detection of variations in levels of ergodicity may *then* be detected as processes of emergence by an observer able to detect the establishment of new levels of coherency by using a suitable cognitive model. Research issues introduced here relate to the possibility of using meta-structures as a *general way* of representing phases.

The subject is of particularly interesting when exploring living matter as matter in a specific phase, like *living*.

We then introduce research issues related to acquisition by matter in the living phase of emergent behavioural properties.

Applications of the approach shortly illustrated above may also be considered as *methodological* for macroscopic phenomena represented by mesoscopic variables identified in a less precise way when considering collective behaviour established in human social systems at the level of description of sociology and economics (see chapter 3), and when inhabiting a structured space like at the level of description of architecture (see chapter 4). In all those case the approach is used not to effectively simulate collective behaviour, but to allow the researcher to conceptually model phenomena by using a mesoscopic level of description and meta-structures possibly useful to hypothesize or assume correspondences effective for influencing emergence of collective behavioural properties. For those cases we introduce some possible mesoscopic variables to be considered.

## 2. APPLICATION OF META-STRUCTURES TO MODELLING PHASES OF MATTER AND, PARTICULARLY, ITS LIVING PHASE.

### 2.1 Phases of matter

In physics a phase is a region of space (a thermodynamic system), where the physical properties of a material are essentially *uniform*, such as having same density.
A phase of a physical system may be defined as a region in the parameter space of the system's thermodynamic variables where the *free energy* is *analytic*.

*Free energy*

In thermodynamics the term free energy relates to a physical variable such that:
- Its changes measure the minimum work the system can do;
- Its minimum values correspond to stable equilibrium states of the system.

The free energy is, for instance, the total amount of energy, used or released during a chemical reaction.
The term relates to the part of the total energy available for *useful work* and not dissipated in *useless work*, as in random thermal motion.
When a system undergoes changes, its free energy decreases.

*Analytic*

In the region in the parameter space of the system's thermodynamic variables the *free energy* can be transformed in an analytic way, i.e., the transforming function is infinitely differentiable and can be described by a Taylor series. In correspondence, we may say that two states of a system are in the *same phase* when they can be transformed into each other with continuity, i.e., without discontinuity amongst their thermodynamic properties.
During a phase transition the free energy is *non-analytic*.
Is *life* a phase of matter like other phases of matter such as the Bose-Einstein condensate, Plasma, Solid, Amorphous Solid, Supersolid, Crystalline solid, Plastic crystal, Liquid Crystals, Magnetic, Superconductive, Liquid, Superfluid, and Gas?

**2.2 Crucial research questions are:**

a) Is a mesoscopic description of phases possible?
b) Consequently, is a meta-structural description of phases possible?
c) Can long-range coherent correlations be modelled by meta-structural properties?
d) Are quantum models of matter *compatible* with meta-structural levels of description?
e) Is it possible to consider collective behaviour as a phase or, better, a process of continuous and coherent changes of phase (Minati, 2008)?

**2.3 Simulating processes of acquisition of emergent properties in artificial systems.**

Is it possible to consider artificial processes able to acquire properties of the kind mentioned above, i.e., suitable to the needs of the composing matter, for instance, electronics?
For instance we could consider the following problem.
How can we make an electronic device to acquire, through continuous, further, processes of emergence occurring within it - ensuring both *generalisation* and *coherence* - behavioural properties corresponding to and even functional for the need to maintain itself working properly?
For instance, we may consider a mobile robot, able to orient, learn, process input, including signals, and perform adaptive strategies. We may assume emergent processes of acquisition of behavioural properties, suitable with its needs such as maintaining its working, occur within it.
How is it possible to make the robot cognitively look for suitable sources of electric power and then be gratified when recharged through a pain-pleasure system without having embedded pre-established representations, i.e., without knowing *in advance*?
Is it possible to design an electronic device able to acquire the emergent property to look for a suitable electric source of power? The problem is introduced in several interesting papers (see, for instance, Grenet and Alexandre, 2008).
The role of the *system pain-pleasure* is very complex because it is not just matter to reward with pleasure or discourage with pain a given action by establishing a feedback to be then substituted through learning.
Complexity relates to key concepts such as:
- Ability to *generalise* behaviours and their correspondence with *levels* of pain or pleasure through learning. This may be accomplished by using available *machine learning techniques*.
- Cognitive transformation of structural needs, such as biological or *electrical* in this case, in emergent behavioural properties, only *after* having been confirmed or regulated by the pain-pleasure system. The hypothesis of pre-established, embedded and sedimented acquired results of ancestral learning processes should be modified with the ability to update and not just replicate. Acquired behaviour is not just *functional*, but *continuously emergent*, *coherent* with the same or equivalent needs in different scenarios. For an electric device the case relates to the search for a suitable electric power socket or from a battery or a solar photoelectric device.

## 2.4 The BioCogniveConverter (BIOCC)

> When matter acquires the property to suffer, desire, and enjoy
> - The becoming turns in thinking-
> Life beyond Maturana and Varela
> Living systems live in a temporary *bubble* of emergent acquired properties,
> i.e., in a temporal reality made of emergent acquired properties.
> From a citation of Fiorella Minati, 1979

Within the emergent process of transformation from properties of matter into behaviour, corresponding to the process of transformation from non-living to living matter (i.e., biochemical emergent matter able to regenerate, reproduce and evolve), it is possible to identify an *ideal general functionality*, namely a *BioCogniveConverter* (BIOCC). BIOCC may be considered as an ideal functionality *acquired* by biochemical systems during the process of emergence. It is presumably established at the upper level of a hierarchy of processes of emergence

*turning* **biological needs such as:**

- The need of adequate food;
- The need of sexual coupling with a *biologically suitable* partner for reproductive reasons;
- The need to protect puppies;

*in* **cognitive evaluations and behaviours such as:**

- Attraction for shapes, colours, odours, sounds, flavours and tactile sensations produced by materiality having biological positive effects and symmetrical repulsion for what having biological negative effects. Biochemical matter then start to feel pain, pleasure, feel scared, cry and smile… There are corresponding attracting effects for further emergent properties such as complex qualities like intelligence, ability to sing, dance, and play music, etc.
- *Targeted* attraction for the opposite sex for the advantage of reproduction by allowing selection depending on species, age, physical conditions, etc. What biologically suitable *looks* cognitively nice and *induce* behaviour.
- Attraction for puppies.

This relates to evolutionary advantages for the acquisition of mind (see, for instance, Popper, 1978), the issues of *qualia* and consciousness of BIOCC in particular. It must be studied the possible correspondence between language and BIOCC. Cognitive sciences, Psychiatry, Genetics, Anthropology, Evolutionism and Psychoanalysis make disciplinary research in this context. We limit ourselves to analyse some possible characteristics of such as ideal functionality:
- It surely has some aspects due to *evolutionary learning* from experience, like drinking, not responsible moreover for all its characteristics. Learning may have adaptive effects.
- It is not *cognitively transmissible* having at the most adaptive effects.
- It allows a kind of autonomous *self-consistence* generating *spirituality* as property compensatory and balancing collateral emergent unwanted effects such as sufferance and delusion. The same ideal functionality applies *by extension* to other different emergent aspects not biologically relevant and generating abstractions such as beauty, good, goodness and *qualia*. Uniqueness of human beings is related to *awareness* of BIOCC, self-usage for development of qualia, knowledge, culture and identification of their limits. BIOCC would be moreover a subsequent, further level of emergence enabling detection and modelling of phenomena otherwise non detectable (*non reductionistic, open* nature of the approach). Phenomena of spirituality assumed as special behaviour in front of a dead companion have been detected, for instance, with chimpanzees and elephants.

- It cognitively *represents* physical states by dreams, visions predictive of physical evolutions (typically pre-death) and sensations.
- Induced behaviours not biologically suitable are considered as *malfunctions* of BIOCC.

BIOCC *allows enduring negative epiphenomena and emergent effects, but also to detect and give symbolic meanings to emergent processes and epiphenomena otherwise completely insignificant.*

- Is it possible to *simulate* such an ideal device?
- Can acquired properties become *autonomous*, i.e., independent from lower physical levels?
- Behaviour may be induced by pre-established bio-chemical process, e.g., sexuality and protection for puppies. Cognitive decisions are actuated by, turned in, mediated by bio-chemical process. The two cases interact and one induces the other. What is a *free* decision?
- What the difference between BIOCC and mind, if any?
- Are we studying BIOCC by using emergent effects of BIOCC itself?

**2.5 Acquisition of properties in living matter**

*"Nature has placed mankind under the governance of two sovereign masters, pain and pleasure."* — Jeremy Bentham

Within the conceptual framework of BIOCC, acquired emergent properties of living matter are assumed to be suitable for biological needs through certain processes.

While it is not clear how biological needs can be *transformed* into behaviour, i.e., the problem of BIOCC, we know of neurological *mechanisms* able to confirm suitability or not of the acquired property. They are all based on pain-pleasure as powerful motivators of behaviour. Recent results showed existence of a common neurobiology for pain and pleasure (Leknes and Tracey, 2008).

We may look for *embedded, innate* levels of meta-structures for mesoscopic variables representing the system pain-pleasure, for instance *the opioid and dopamine systems* and antinociceptive effects.

They may be considered to encourage or discourage, rather than to cognitively support design or recognition, of activities suitable or non-suitable to the needs of biological matter to remain in the living *phase*. This may always be combined with further learning processes.

*Embedded, innate* levels of meta-structures for mesoscopic variables representing the system pain-pleasure may have interesting relationships with the concept of consciousness itself and the emergence of mind.

*Continuity* between properties of non-living and living matter (Rasmussen *et al.*, 2003; 2004) can relate to their meta-structural composition (Minati, 2008a, pp.107-110).

Can structure acquired by matter in the living phase (?) be modelled as produced by meta-structural properties as for protocells (Serra *et. al.*, 2007; 2008a; 2008b)? Which relations with a quantistic description of living matter (Vitiello, 2004; Vitiello, in press)? See, for instance, the project *Physics of Living Matter* in the web resources.

Where do such meta-structural constraints come from?

**Meta-structural constraints would be considered *intrinsic* to the matter. In such a way, living is just one of the possible phases of matter in a comprehensive coherent scenario without discontinuities.**

Examples of research topics are:
- Are quantum models of matter in the living phase *compatible* with meta-structural constraints at mesoscopic levels? This point relates, in particular, to *quantum correlations* (entanglement), see, for instance, (Vedral *et al.*, 1997; Vedral and Plenio, 1998).

*From property of matter to behavioural decisions: can Meta-structural Systems Analysis help?*
- Conversion of bio-chemical constraints of living systems into behaviour
- Conversion of properties of non-living matter into decisions
- Qualia. (Minati, 2008a, pp.107-110).

# 3. APPLICATION OF META-STRUCTURES FOR MODELLING PROCESSES OF ACQUISITION OF EMERGENT PROPERTIES IN SOCIAL SYSTEMS.

> "The great basic thought that the world is not to be comprehended as a complex of readymade *things*, but as a complex of *processes*, in which the things apparently stable no less than their mind images in our heads, the concepts, go through an uninterrupted change of coming into being and passing away, in which, in spite of all seeming accidentally and of all temporary retrogression, a progressive development asserts itself in the end — this great fundamental thought has, especially since the time of Hegel, so thoroughly permeated ordinary consciousness that in this generality it is now scarcely ever contradicted."
>
> Friedrich Engels
> In *Ludwig Feuerbach and the End of Classical German Philosophy*. Part 4: Marx
> First Published: 1886, in Die Neue Zeit;
> Source: Progress Publishers edition;
> Translated: by Progress Publishers in 1946;

On the basis of the theoretical approach for modelling processes of emergence as in chapter 1 the purpose is to consider meta-structures as a suitable methodological approach for modelling processes of emergence in social systems (Keith, 2005).

Current interdisciplinary approaches are assumed be applications, i.e., *transpositions*, of models introduced in other disciplines, usually physics, changing the *meaning* of variables.

Social scientists criticize such an approach as *reductionistic* because the whole complexity of human behaviour should not be reduced to single variables as for physics. Moreover, the approach is based on the so-called *homogeneous hypothesis* when elements are assumed to be *indistinguishable*. In this case elements are assumed to be particles and their interaction may be modelled by mathematical equations and often by very simple rules. An example is given by gases consisting of particles and adopting systemic properties such as pressure and temperature. The hypothesis may apply even when interacting elements are autonomous systems, i.e., provided with cognitive systems, all being considered as equal in a simplified, reductive, way. This is, for example, the case for models based on agents interacting according to a few, simple rules (e.g., eco-systems and markets).

A different, more realistic approach is based upon considering elements as being different and distinguishable, as in the so-called *heterogeneous assumption*. In this case each element interacts in different, emergent ways. This is the typical case of autonomous agents *processing* interactions and not simply reacting. Here, the processing is performed by the cognitive system and the result is non-deterministic. A typical example is given by families of human beings. Human beings establish systems, in this case families, assuming sociological properties different from those of its components, such as decisions emerging from discussions, i.e., interactions, regarding educational choices for children and economic behaviour. In some cases the cognitive system is so elementary that it is possible to simplify, by adopting a suitable particle representation, as for swarming and flocking modelled by assuming that elements react according to very simple rules.

Moreover, approaches for suitable generalised modelling of such very complex phenomena are still not available. This is due to the lack of general theories of processes of emergence. Attempts have been made, such as for Multiple Systems and Collective Beings (Minati and Pessa 2006).

The project introduced in this chapter will explore suitability of meta-structural properties to model processes of emergence in social systems.

**3.1 Modelling processes of acquisition of emergent properties in social systems.**

We introduce a possible *methodological way* to use meta-structural analysis as research project having the purpose to model acquisition of emergent properties in human social systems.

The research is based on modelling acquisition of emergent collective behavioural properties in human social systems by using mesoscopic variables.

In this framework we will consider *cognitive self-acquired constraints* used by agents to interact as *boundary conditions* or operators able to induce acquisition and then maintain emergent properties in human social systems.

We hypothesize that *respect of boundary conditions* and *usage of operator*s by agents when interacting may be suitably represented by meta-structures.

In the project outlined in chapter 4 and related to architecture, constraints considered will be *material one*, e.g., geometrical and topological, rather than cognitive as in the case we are now dealing with.

Examples of cognitive behavioural constraints self-acquired and then structured, thank to sociological sedimentation processes, by social systems are:

- Rites (religious and social, as *sociological needs* of living systems provided with complex cognitive systems able to represent and perform symbolic and sub-symbolic processing);
- Language: the *Sapir-Whorf hypothesis* revisited. Consent Manipulation (Minati and Magliocca, 2008; Minati, 2006);
- Music, meta-structure of thinking (going *beyond* onomatopoeic analogy and as musical accompaniment for *explicit* text, Minati, 2002a);
- Habits representing and inducing behaviour;
- Traditions (respect for) representing and inducing behaviour;
- Fashion representing and inducing behaviour;
- Style representing and inducing behaviour;
- Ethics as *social software* (Minati, 2002b);

Such cognitive behavioural constraints are of great interest for modelling behaviour acquired by human social systems, because human beings use complex cognitive systems to create and use representations in symbolic and sub-symbolic ways. Human social systems are considered made emergent by the interaction of agents respecting such cognitive constraints and using them as operators.

We consider the possibility to represent the processes outlined above by using meta-structures, based, for instance, on the following mesoscopic variables:

- How many people repeat the same behavioural action over time? Example of behavioural actions in this case may be: driving a car; using the train, bus, and airplane; buying a service or a good; watching TV; eating; sitting; telephoning; etc.

Other more specific cases may be:

- How many people dress in the same way over time?
- How many people use the same linguistic expression over time?
- How many people hear the same music over time?

Meta-elements may relate to the time spent by agents in a specific state and allowing for an ergodic analysis. Besides it is possible to also consider specific corresponding values of parameters defining mesoscopic variables, like the specific car, service, good, TV program, etc.

Meta-structures are expected to represent the collective behaviour assumed, but they are also expected as possible way to induce and modify the collective behaviour acquired.

We mention as the *superstructure* in Marx's theory of history may be revisited as an acquired emergent meta-structure. This may help to overpass the strict causal relation between economic base and ideological superstructure, as in Marx's 1859 text, *A Contribution to a Critique of Political Economy* and more adequately deal with the timeliness *necessary* coherence between structure and superstructure to allow equilibrium in social systems. The novelty in the project we are considering lies in the way by which they correspond, i.e. *non-linear, even non-functional, but rather emergent, sub-symbolic* (the real innovation compared to Marx). Marx's reductionism consists of the *correspondence* between structure and superstructure, intended to *functionally* represent the structure without considering superstructure as a property acquired through processes of emergence occurring in the structure and processes of emergence occurring in the superstructure itself. The observer should use different *simultaneous* levels of description, their interactions and multiple modelling in a contructivist way like when using *logical openness* and the concepts of Multiple Systems and Collective Beings (Licata, 2008a; 2008b; Minati *et. al.*, 1996; 1998; Minati and Pessa, 2006).

*Meta-structures may be intended, in the computational view, as establishing social software for social systems* (Minati and Pessa, 2006).

### 3.2 The biological device assuming behaviour and using a cognitive system: the *BioCogniveConverter* (BIOCC), see project No. 2.

Examples of research topics are:
- Design *freedom*? Consciousness and meta-structural freedom.
- Does biological matter need it?
- Is consciousness the upper level of logical openness?
- Life as a phase of matter
- Death as a phase transition of living matter
- Biological devices able to perform the phase transition by which acquired emergent properties, like mind, can be sustained by other processes no longer based on living biological matter.

### 4. ARCHITECTURE AS THE SELF-DESIGN OF META-STRUCTURES (BOUNDARY CONDITIONS) TO INDUCE AND MAINTAIN, THROUGH STRUCTURING, EMERGENT PROPERTIES IN HUMAN SOCIAL SYSTEMS.

We introduce a possible *methodological way* to use meta-structural analysis as research project having the purpose to model acquisition of emergent properties in human social systems.

The research is based on modelling acquisition of emergent collective behavioural properties in human social systems by using mesoscopic variables.

In this framework based on considering the search for meta-structural properties as a methodology able to display meta-structures and make them usable to induce and change collective behavioural acquired properties in social systems, we will consider concrete structures of living space made by architecture as *boundary conditions* or operators able to induce acquisition and then maintain emergent properties in human social systems.

We hypothesize that *respect and usage of boundary conditions* by agents when interacting may be suitably represented by meta-structures.

In a systemic approach a proper definition for architecture could be "the set of human artefacts and signs that establish and denote mankind's settlements", inspired to the one introduced by William Morris (1834-1896): *all the signs that mankind leaves on the Earth except pure desert* (*The prospect of architecture in civilization*, London, 1881). In this conceptual framework, to represent, analyze, understand, design and act in the built environment we must resort to a great number of different disciplines in a systemic way (Di Battista, 2008; Di Battista *et al.*, 2006; Di Battista, 2005).

Since human settlements are the product of human societies, they are mostly built up and developed **by a huge number of interacting unconscious acts** during a long time, rather than by purposely single *designed acts*; such a vision generates the idea of an *implicit project* that relies upon the systemic approach, like in:

> "Architecture organizes and represents the settlement system; it interprets, materializes, interacts with and confirms the references of cognitive systems, and projects (foresees) and builds "coherent occurrences" (steadiness, confirmation) and "incoherent occurrences" (emergence) in the settlement itself. Architecture operates in the interactions between mankind and natural environment with "coherent actions" (communication; consistent changes; confirmation of symbols and meaning) and "incoherent actions" (casual changes, inconsistent changes, new symbols and meanings). Coherent actions are usually controlled by rules and laws that guarantee stability to the system (conditions of identity and acknowledged values); incoherent actions generally derive from a break in the cognitive references (breaking the paradigm) or from the action of "implicit projects". These are the result of multiple actions by different subjects who operate all together without any or with very weak connections and have different – sometimes conflicting – interests, knowledge, codes, objectives. Implicit projects always act in the crack and gaps of a rule system; they often succeed, according to the freedom allowed by the settlement system.

Perhaps, the possible virtuous connections of this project, in its probable ways of organization and representation, could identify, today, the boundaries of architecture that, with or without architects, encompass "the whole of artefacts and signs that establish and define the human settlement". "(Di Battista, 2006, pp. 8-9).

and

"In the open system of the built environment and in the continuous flow of human settlements that inhabit places, there are many reasons, emotions, needs, all of which are constantly operating everywhere in order to transform, preserve, infill, promote or remove things. These intentional actions, every day, change and/or confirm the different levels of our landscape and built environment. This flow records the continuous variation of the complex connexions between people and places. This flow records the continuous variation of the complex connexions between people and places. This flow represents and produces the *implicit project* that all built environments carry out to update uses, values, conditions and meaning of their places. This project is implicit because it is self-generated by the random summation of many different and distinct needs and intentions, continuously carried out by undefined and changing subjects. It gets carried through in a totally unpredictable way – as it comes to goals, time, conditions and outcomes.

It is this project anyway, by chaotic summations which are nevertheless continuous over time, that transforms and/or preserves all built environments.

No single project, either modern or contemporary, has ever been and will ever be so powerful as to direct the physical effects and the meanings brought about by the *implicit project*." (Di Battista, 2008, pp. 45-46).

### 4.1 The project *self-architecture*

The concept of boundary conditions is used in mathematics. Dealing with differential equations, the *boundary value problem* is given by a differential equation *and* a set of additional restraints, called the boundary conditions.

We will generalize the concept of boundary condition to the degrees of freedom or constraints given by structures, e.g., geometrical and topological properties of *living space* as shaped by architectural design, and interacting agents establishing collective behaviour.

Meta-elements are introduced as *sets* of time-ordered values in a discrete temporal representation adopted by suitable mesoscopic state variables describing *global, collective aspects* of the system under study. The properties of meta-elements are expected to represent *aspects* of a more general and, consequently, more suitable level of description for collective behavioural phenomena.

Meta-structures are assumed to represent the properties of meta-elements and their possible relationships. Meta-structures may be intended as degrees of freedom of elements at a more general level of description, i.e., those of meta-elements, able to *indirectly influence* the behaviour of agents described at a lower level of description and producing collective phenomena by non-linearly *completing* a partial structure.

The general aims of the project are to identify, use and suitably design architectural projects as meta-structural designs not only having *functional consequences* for social systems behaviour, but emergent effects. More specifically, meta-structures in architectural designs will be considered as

   a) Expression, materialisation of emergent properties (at different levels of descriptions such as sociological, religious, artistic, military, economic and scientific) acquired by social systems. In this case meta-structures have the purpose of making emergent, acquired properties *structural* and no longer emergent;

and, conversely:

   b) As a way to influence, or induce processes of emergence within social systems in such a way as to lead to the acquisition of some specific properties, coherent with current evolutionary processes at different levels of description such as sociological, religious, artistic, military, economic and scientific.

Examples of case a) are architectures of dwellings intended first as the materialization of kinds of housing and then their inducement; architectures of hospitals intended first as the materialization of therapeutic and medical approaches and then their inducement; and architectures of schools intended first as the materialization of ways of considering knowledge, i.e., disciplinary fragmentation, and then its inducement. Other examples are the shape of roads influencing traffic and the number of entrances-exits or the surface area of a flat influencing its inhabitants' social behaviour.

Other examples of boundary conditions affecting collective behaviours and inducing emergence of social behavioural properties are:
1. Number of entries and exits for flats;
2. Central role assumed by some functional areas in flats, like the kitchen coming from the age where it was the warmest place;
3. Available living surface inducing residence for singles or families;
4. Shapes of roads inducing properties of traffic;
5. Number of baths per inhabitants;
6. Form walls and topology usually fit with roles;
7. Availability of sidewalks inducing or preventing pedestrian traffic;
8. Lighting making living styles possible;
9. Stairs, e.g., stairs with one handrail and with two handrails; width allowing usages;
10. Internal facilities (private, inducing *consuming*, e.g. shopping centres) rather than external (public)

Moreover, architecture does not only materialise and transform acquired emergent properties of social systems into structural constraints, but it also induces new emergent properties when introducing innovative ways of structuring space as in case 2). Examples are vertical constructions, e.g., skyscrapers, underground constructions and cities.

The project has been named *self-architecture* (Minati and Collen, 2008; Minati, 2008g).

Therefore, a systemic approach to architecture and its understanding entails several disciplines including engineering, design, psychology, anthropology, sociology, economics, and the whole *science of complexity* made, in its turn, by inter-disciplinary models and studies in physics, mathematics, cognitive sciences, artificial life, artificial intelligence, meteorology, earth sciences, chemistry and biology.

## 4.2 Research topics

Some research topics are:
1. Is it possible to identify meta-structural properties affecting emergence of collective behaviour performed by inhabitant agents?
2. Consequent ethical problems for the designer of structures as constraints able to affects processes of acquisitions of new emergent properties in social systems, the architect in this case.
3. Study of the *coherence* and *continuity* between disciplinary aspects of social systems, mutually interacting and *represented* one into the other such as the ones expresses by literature, architecture, language, music, painting, religion and science.
4. Landscape as result of global emergent effects at any scale.
5. Relationships between scaling.
6. Cities as Collective Beings (Batty, 2008; Minati, 2008f).
7. Pre- and Post-Occupancy and Building Performance Evaluations (POE, BPE) as Meta-Structural analysis.
8. Models and simulations of emergent effects in architecture, by using the software for simulation and search for meta-structures to be released and *free downloadable*.
9. How does the phenomenon of induction work in architecture for meta-structural properties?
10. Emergent phenomena in architecture.

## 4.3 Modelling processes of acquisition of emergent properties in social systems through architecture

Examples of mesocopic variables in social systems inhabiting a structured space assumed influencing acquisition of emergent behavioural properties are:
- How many people are seated over time?
- How many people are using stairs over time?
- How many people are walking over time?
- How many people are using an elevator over time?
- How many people are using artificial light over time?
- How many people are entering-outgoing over time?

Mesoscopic variables may be specified by considering details such as spatial dimensions of actions. Meta-elements may relate to the time spent by agents in a specific state and allowing for an ergodic analysis. Besides it is possible to also consider specific corresponding values of parameters defining mesoscopic variables.

Meta-structures are expected to represent the collective behaviour assumed as corresponding to material structures of living space intended as constrains-boundary conditions, operators to pursuit an action like reaching a place, but they are also expected as possible way to induce and modify the collective behaviour acquired.

Meta-structures will also *represent* the kind of inhabitant by assuming different properties for homogeneous agents like children in kindergartens, old people in homes for the elderly, patients in hospitals, students in schools and barracks for soldiers.

## CONCLUSIONS

We presented a short overview on the usage of meta-structures, introduced in the literature, to model acquisition of behavioural properties in social systems of agents. We mentioned some approaches based on considering ergodical properties and software tools under development for simulation.

The purpose of the paper is to introduce the reader to some possible interdisciplinary applications of the approach suitable to establish corresponding research projects presented in the chapters.

## WEB RESOURCES